# ICT Convergence in Internet of Things – The Birth of Smart Factories *(A Technical Note)*

*Mahmood Adnan, Hushairi Zen*

*Abstract* – **Over the past decade, most factories across developed parts of the world employ a varying amount of the manufacturing technologies including autonomous robots, RFID (radio frequency identification) technology, NCs (numerically controlled machines), wireless sensor networks embedded with specialized computerized softwares for sophisticated product designs, engineering analysis, and remote control of machinery, etc. The ultimate aim of these all dramatic developments in manufacturing sector is thus to achieve aspects such as shorter innovation / product life cycles and raising overall productivity via efficiently handling complex interactions among the various stages (functions, departments) of a production line. The notion, *Factory of the Future,* is an unpredictable heaven of efficaciousness, wherein, issues such as the flaws and downtime would be issues of the long forgotten age. This technical note thus provides an overview of this awesome revolution waiting to be soon realized in the manufacturing sector.**

*Index Terms* – **Smart Factories, Fourth Industrial Revolution, Internet of Things, Ubiquitous Computing.**

## I. INTRODUCTION

WE often perceive Factories as dirty and noisy production amenities over - crowded with manpower and machinery incessantly executing same tasks again and again with abundant flaws and downtime. But now, Factories are about to get really smarter. We're rapidly heading towards fourth industrialization revolution, wherein, everyday machines that bakes sandwiches to blending of coffee and sophisticated machines manufacturing cellular phones and automobiles are about to get revolutionized. 'Smart Factories or Factories of the Future' (terminologies used interchangeably) are said to be employing a notion of *'Internet of Things'* for operating complex manufacturing processes via hundreds and thousands of computers depending on the size of a particular industry.

Global competition for Smart Manufacturing is undoubtedly ravenous and Germany has taken this race with introduction of the first promising standard, i.e. Industry 4.0; and Government has accordingly invested amount to the tune of approximately €500m for materializing this technology [1, 2]. DFKI – German Artificial Intelligence Research Centre is now regarded as one of the acclaimed research centres worldwide for production of Smart Factories technology. European Commission's worth $2 billion project, *Factories of Future - Public Private Partnership* also aims to develop footprints / trails of Smart Manufacturing in European Union. Steps are now being taken to create a global Standard in certain developed parts of the world such as USA, China, Korea, and Japan while the rest of the word still appears at quite a silence.

## II. SMART FACTORIES – DEFINITION & CHALLENGES

Although there is no universal definition that defines aspects of Smart Factories, several research groups have their prophetic viewpoints. Smart Factories are actually envisaged on notion of *Internet of Things*, referred in the industrial sector as *'Industrial Internet of Things (IIoT)'*. The main intent is to empower all the assets / physical things on the shop floors, assembly lines, and batches with digital or virtual voice, which in turn enables them to communicate some sort of information about themselves (i.e. their status – what are they, where are they, their conditions like temperature, and hence so forth). If employed precisely, these interlinked devices from a convergence point between physical and digital world would empower the today's industrial systems to transform into much smarter / intelligent systems [2].

Since the data from the shop floors (machine shop), assembly lines, and batches gets connected to a Cloud and is readily made observable in real time – successful companies can thus achieve shorter innovation and product life cycles, can add value to their offerings through faster time to the market, efficacious product personalization, transparency in their operations, well-informed processes, etc. The real challenges however lies in achieving its design principles including (but not limited to) *Interoperability, Virtualization, Real-Time Capability, Decentralization, Service Orientation, and Modularity, etc.*

## CONCLUSION

The Smart Factory is believed to be a proclaimer of the fourth industrial revolution, an important swing in the operating norms that are expected to alter the mode via which the manufacturing companies today operates. It is just the matter of time; the world would soon see either a big IT-Multis, any visionary automation leader, or a small aspiring start-up to materialize this sphere.

## REFERENCES

[1] A. Radziwon, A. Bilberg, M. Bogers, and E. S. Madsen (2014), "The Smart Factory, Exploring Adaptive & Flexible Manufacturing Solutions", Procedia Engineering, Vol. 69, pp. 1184-1190.

[2] S. Wang, J. Wan, D. Zhang, D. Li, C. Zhang (2016), "Towards Smart Factory for Industry 4.0: A Self-organized Multi-agent System with Big Data Based Feedback & Coordination", Computer Networks (Elsevier), [doi:10.1016/j.comnet.2015.12.017].

Mahmood Adnan is associated with the Faculty of Engineering, Universiti Malaysia Sarawak in the capacity of Postgraduate Researcher.
Hushairi Zen is Deputy Dean (Industry and Community Engagement) with the Faculty of Engineering, Universiti Malaysia Sarawak.